\documentclass[twocolumn,
runinaddress,
preprintnumbers,
noeprint,
aps,amsmath,amssymb,
]{revtex4-2}

\usepackage[utf8]{inputenc}

\usepackage{amsmath}
\usepackage{amssymb}
\usepackage{graphicx}
\usepackage{esint}
\usepackage{hyperref}
\usepackage{multirow}
\usepackage{color}
\usepackage[makeroom]{cancel}
\usepackage[normalem]{ulem}

\begin{document}

\title{Observation of two zeros of the real amplitude in pp scattering at LHC energies}

\author{Anderson Kendi Kohara}
\email{kohara@agh.edu.pl}

\affiliation{ AGH University Of Science and Technology, Faculty of Physics and Applied Computer Science, \\ Mickiewicza 30, 30-059 Krak\'ow, Poland
}


\begin{abstract}

  Elastic scattering of charged hadrons is described by the combination of nuclear and Coulomb amplitudes. It is well know that at the very forward range the nuclear real and Coulomb parts interplay a crucial role in the determination of the magnitude of the real part at $|t|=0$. However, beyond $|t|=0$ the real and imaginary nuclear amplitudes have different $t$ dependencies and we show that at LHC energies the zeros formed by the combination $T_C(t)+T_R^N(s,t)=0$ in pp process can be potentially observed when the background due to the imaginary part is removed. This observation constrains the real part at this forward range. 
\end{abstract}

\maketitle

\section{Introduction}

The complex amplitudes in quantum mechanics are not a direct measurable quantity, living in the abstract Hilbert space, while the measurable absolute square gives the probability of finding particles according to certain distribution. In elastic scattering of charged hadrons, the strong (complex valued) and  electromagnetic (pure real) forces interplay and the interference between these quantities can be in principle observed in experiments, constraining the amplitudes. Also,  some theorems associated with unitarity and analiticity of the nuclear amplitudes constrain the behaviour of these objects.

In the very forward range, the  Coulomb amplitude drops down fast, and the differential cross section becomes  dominated by the nuclear parts. The optical theorem, which extrapolates the imaginary part at $|t|=0$ relates the magnitude of this amplitude to the total cross section.
In 1970, before the experimental results pointing to the rise of pp and p\=p total cross sections, Cheng and Wu, based on a massive electrodynamics, predicted that $\sigma$ should saturate the Froissart bound at infinity energies \cite{PhysRevLett.24.1456}. For an increasing cross section such as $\sigma \sim \mathcal{C}\log^2 s$ the dispersion relations predict the parameter $\rho \sim \pi/\log s$, where $\rho$ is the ratio of the real and imaginary amplitudes as $|t|=0$. 
Recently, Martin and Wu 
formally proved \cite{Martin_2018} that at large energies, if the total cross section goes monotonically to infinity at infinity energies, the real amplitude is positive in the forward direction. In addition, it was also proved by Martin that if the differential cross sections for crossing symmetric processes $d\sigma/dt({\rm ab \to ab})$ and $d\sigma/dt({\rm a\bar b \to a\bar b})$ tend to zero for $s\to \infty$ in a strip $0 < |t| < |\widetilde{t}|$ where $|\widetilde{t}|$ is arbitrarily small and if $\sigma({\rm ab \to ab})$ and/or $\sigma({\rm a\bar{b} \to a\bar{b}})$ tend to infinity for $s \to \infty$  the real part cannot have a constant sign \cite{Martin1997ATO}, which means that the real part must cross zero at some $|t_R|$ within the diffractive cone. This zero of the real nuclear part ($|t_R|$) is dubbed Martin's zero.  In the 1970s it was shown that for high energies, in the geometric scaling regime, the real nuclear part of a crossing symmetric amplitude in the forward range  has a zero approximately at $|t_R| \simeq 1/\log s$ \cite{Martin:880639}, i.e, it is shrinking with increasing energy.

\begin{figure}
\includegraphics[ scale=0.65]{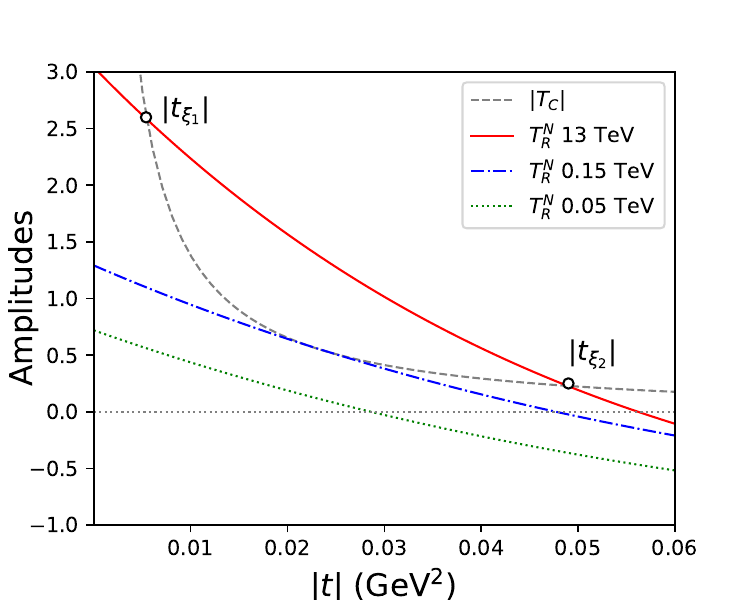}
\caption{We show the evolution of the  nuclear real amplitude from ISR to LHC energies \cite{Kohara:2019qoq} and the absolute value of the Coulomb amplitude. We  observe that around some critical energy $\sqrt{s_c}=0.15$  TeV there is some $|t|$ where $T_R^N=|T_C|$. Beyond this energy $T_R^N$ becomes larger than $|T_C|$ within some region $|t_{\xi_1}|<|t|<|t_{\xi_2}|$.}
\label{amplitudes-increasing}
\end{figure}

Extending the ideas of  crossing symmetric amplitudes in the forward range, a  phenomenological model for pp and p\=p amplitudes was proposed describing the scattering data from ISR to LHC energies and the analytical form for $|t_R(s)|$ was suggested \cite{Kohara:2019qoq}. The model satisfies dispersion relations since it connects the real and imaginary parts analytically. According to the model, in the ISR energies the real nuclear  amplitude for both pp and p\=p is always smaller than the absolute value of the Coulomb amplitude. However, when the energy increases, as can be seen in Fig. \ref{amplitudes-increasing}, eventually the real nuclear part equates $|T_C|$ at some $|t|$ and for larger energies, say, the LHC range, $T_R^N(s,t) > |T_C(t)|$ in some region $0<|t|<|t_R|$.

Using the above ingredients, in the present work we show that there might exist 
some critical energy $s_c$ such that beyond it $T_R^N(s,t)>|T_C(t)|$ for some region $0<|t|< |t_R|$ and if this condition is satisfied, we prove that in the pp case, the sum of the pure real and Coulomb amplitudes has two zeros within the diffractive cone.   We then suggest how these zeros in pp scattering could be extracted from the data and where they are expected to be observed in the LHC range. 
\section{Emergence of Two Zeros in the Real Amplitude}

We wish to show that the existence of a region  where possibly $T_R^N(s,t)>|T_C(t)|$ is not a particular feature of a model  and in principle can be observed in different models with independent real and imaginary nuclear amplitudes. For this purpose, let's consider a moving point $|t_m(s)|$  in the region $0<|t_m|<|t_R|$ defined as 
\begin{eqnarray}
|t_m(s)|\equiv|t_R(s)|-|\eta(s)|=\big|\frac{t_{R0}}{\log s}\big|-\big|\frac{\eta_0}{\log s}\big| ~ , 
\label{zero_martin}
\end{eqnarray}
 such that $ ~~ 0<|\eta|<|t_R|$, where $t_{R0}$ is some constant determined from the phenomenology and $\eta_0$ is some arbitrary constant. Since $|t_R|$ is shrinking as $\sim 1/\log s$ it is natural to expect that $|\eta| \sim 1/\log s$  also shrinks, in order to satisfy $0<|\eta|<|t_R|$.
 On the other hand, since the Coulomb amplitude for pp is $T_C(t)\sim - \alpha/|t|\,\,$ it means that at $|t_m(s)|$  we have
 \begin{eqnarray}
  &&T_C(t_m)\sim - \frac{\alpha}{|t_R|-|\eta|}=-\frac{\alpha\log s}{|t_{R0}|-|\eta_0|}~.
 \label{Coulomb_at_tc}
 \end{eqnarray}
 As discussed in the introduction, for large energies, if the total cross section behaves  as
 $\sigma \sim \mathcal{C}\log^2 s$, since $\sigma\propto T_I^N(s,t=0)$, the real amplitude at the origin behaves as
 \begin{eqnarray}
 T_R^N(s,t=0)\sim \mathcal{C}\pi\log s ~ .    \label{real_dr}
 \end{eqnarray}
  Very close to the origin the real amplitude falls as an exponential, much slower than the Coulomb part. In this sense at $|t_m|$ the magnitude of the real part is slowly varying  and we could safely compare the real part at $t=0$ i.e, Eq.(\ref{real_dr})
  with the absolute value of Eq. (\ref{Coulomb_at_tc}), and if $\,\mathcal{C}\,\pi \gtrsim \alpha/(|t_{R0}|-|\eta_0|)$ we have $T_R^N(s,t_m)>|T_C(t_m)|$, i.e, a dominance of the real nuclear amplitude over the Coulomb part at $|t_m|$ where $0<|t_m|<|t_R|$. The existence of such dominance is present in several models of elastic scattering. Although not explicitly mentioned in our previous works, where we study a broad $t$ range, this dominance also occurs \cite{kohara:2014, Kakkad_2022}.

Starting from the region where $T_R^N(s,t)>|T_C(t)|$, when $|t|$ approximates to $|t|=0$, the Coulomb amplitude decreases very fast to $-\infty$. Therefore,  we have the inequality,
\begin{eqnarray}
T_R^N(s,t)<|T_C(t)|, ~~ |t|\to 0~.
    \label{inequality_TR_let_TC}
\end{eqnarray}
This means that for some non zero value  $|t|=|t_{\xi_1}|<|t_{R}|$, the real amplitude crosses  the absolute value of the Coulomb amplitude,
\begin{eqnarray}
T_R^N(s,t_{\xi_1})=|T_C(t_{\xi_1})|, ~~ 0<|t_{\xi_1}| < |t_{R}|~.
    \label{TR_equal_TC_at_txi1}
\end{eqnarray}
  
On the other hand, as a consequence of the existence of a zero $|t_R|$, the real part of the nuclear amplitude $T_R^N$ decreases monotonically as function of $|t|$  from the origin  towards $|t_R|$.
Thus, after crossing the region where 
\begin{eqnarray}
T_R^N(s,t)>|T_C(t)|, ~~ |t_{\xi_1}|<|t|<|t_R|  ~,
    \label{inequality_TR_get_TC}
\end{eqnarray}
since $|T_C(t)|$ will never cross zero, eventually $T_R^N(s,t)$ will reach again the absolute value of the Coulomb amplitude for some value $|t|=|t_{\xi_2}|$ 
\begin{eqnarray}
T_R^N(s,t_{\xi_2})=|T_C(t_{\xi_2})|, ~~ |t_{\xi_1}|<|t_{\xi_2}| < |t_R|~. \label{TR_equal_TC_at_txi2}
\end{eqnarray}

\begin{figure}
\includegraphics[ scale=0.65]{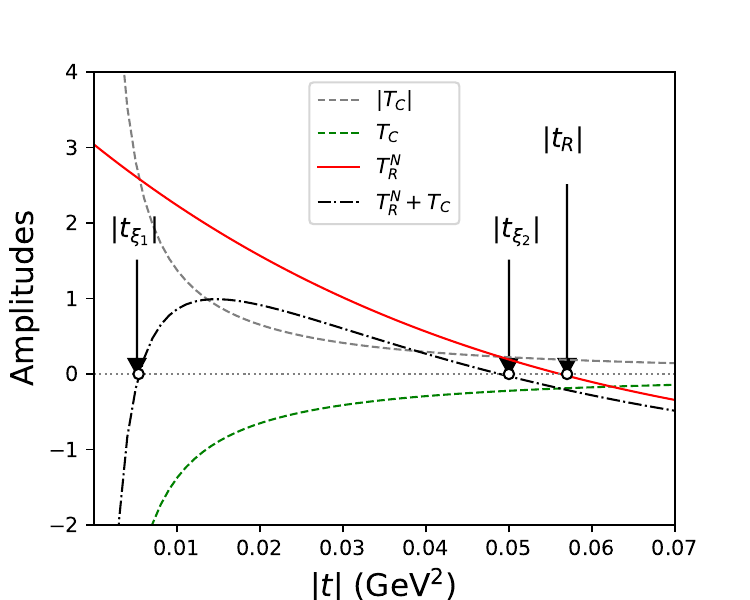}
\caption{We show the nuclear real, the Coulomb, the absolute value of the Coulomb and the combined nuclear real and Coulomb amplitudes. Within the region $|t_{\xi_1}|<|t|<|t_{\xi_2}|$ we represent $T_R^N > |T_C|$ and as we prove, the combined $T_R^N+T_C$ has two zeros, one at $|t_{\xi_1}|$ and the other at $|t_{\xi_2}|$. }
\label{scattering-amplitudes}
\end{figure}

The above arguments can be summarized as follows: Let $T_R(s,t)$ be the real part of the sum of the nuclear and Coulomb pp amplitudes,
\begin{eqnarray}
T_R(s,t)\equiv T_R^N(s,t) + T_C(s)~,
    \label{total_real_amplitude}
\end{eqnarray}
then, for $s$ large, if $T_R^N(s,t)>|T_C(t)|$ in a region $0<|t|<|t_R|$ then $T_R(s,t)$ has two zeros,
\begin{eqnarray}
T_R(s,t_{\xi_1})=T_R(s,t_{\xi_2})=0, ~~ 0<|t_{\xi_1}|<|t_{\xi_2}|<|t_{R}| ~.
    \label{h_and_two_zeros}
\end{eqnarray}
 In Fig.\ref{scattering-amplitudes} we represent the above proposition, showing the situation where $T_R$ has two zeros. Despite the simplicity  on the arguments about the existence of the first zero in pp scattering, it is not completely clear yet for which energy the first zero emerges. The possibility of the existence of the first zero was pointed out previously in Ref. \cite{Oleg:2001}.

For the sake of simplicity in the above arguments we neglect the effects of the hadronic electromagnetic form factor and the relative Coulomb phase, but including these ingredients, qualitatively the conclusions should remain the same.  In our previous works we already discussed the emergence of the zeros  $|t_{\xi_1}|$ and $|t_{\xi_2}|$ in the LHC energies \cite{Kohara_2017,Kohara_2019} and also we test the effects of Coulomb phase.  

In this letter we wish to point out that  unlike in the ISR energies, the LHC range shows some evidence for the existence of these zeros.

\section{Amplitudes and observables}
\vspace{5pt}
The differential cross section is formally written as
\begin{eqnarray}
\frac{d\sigma}{dt}=\frac{1}{16\pi(\hbar c)^2}|T^N(s,t)+T_C(t)e^{i\alpha\Phi(s,t)}|^2
\label{diff_cross_formal}
\end{eqnarray}
where $T^N(s,t)$ is the complex nuclear amplitude and $\Phi(s,t)$ is the relative Coulomb phase. 

The Coulomb amplitude is standard and is given by
\begin{eqnarray}
T_C(t)=-\frac{2\,\alpha\, G(t)^2 }{|t|}=-\frac{2\,\alpha}{|t|}\Big(\frac{\Lambda^2}{\Lambda^2+|t|}\Big)^4
\label{Coulomb}
\end{eqnarray}
where $\alpha=1/137$ is the fine structure constant, $G(t)$ is the proton electromagnetic form factor and $\Lambda^2=0.71$ GeV$^2$ is a momentum scale.

The nuclear amplitude depends on the model, but in the forward range, since the differential cross section is decreasing approximately as an exponential it is natural to parametrize the nuclear amplitudes with exponential forms.
However, it is now clear that the real and imaginary amplitudes must have different $t$ dependencies and since the imaginary part is much larger than the real amplitude in the forward range in order to observe such subtle effects i.e., the existence of the two zeros, one needs to remove the background due to the imaginary amplitude. 

In the LHC analysis, in order to show some non trivial behaviour in the data, it has been common to present the very forward region by subtracting and then dividing the differential cross section from a reference, defined simply such as $Ref(t) = A\,e^{-B\,|t|}$, where $A$ and $B$ are obtained by fitting the very forward data. As a result, the subtracted quantity shows some non-linear behaviour  as function of $t$. 

\begin{figure}
\includegraphics[ scale=0.64]{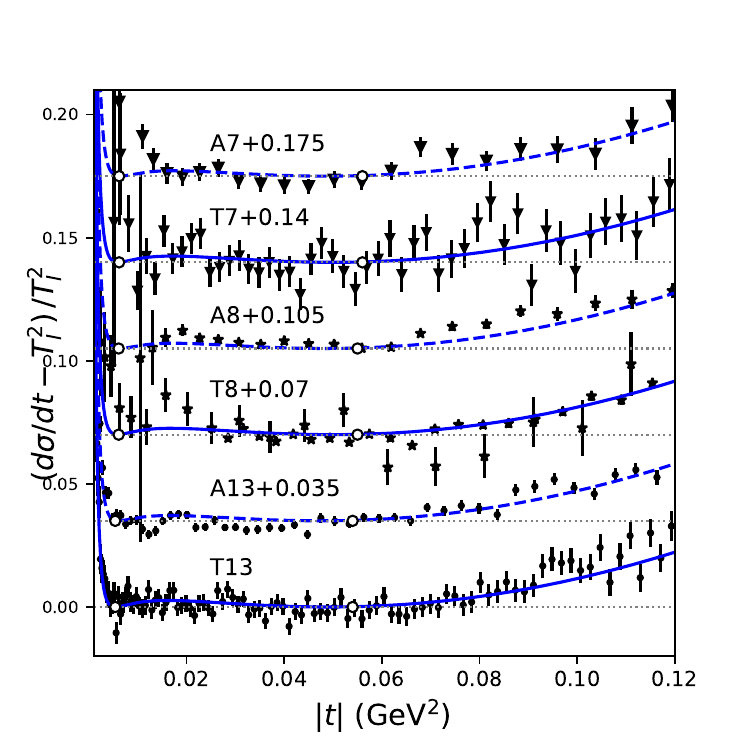}
\caption{Subtracted Totem (T) and Atlas (A) data sets for 7, 8 and 13 TeV up to $|t|=0.12$ GeV$^2$ compared with lines calculated from the model proposed in Ref.\cite{Kohara:2019qoq}. For each experiment we add a factor to  unstack  the data sets. The open circles represent the positions of the zeros $|t_{\xi_1}|$ and $|t_{\xi_2}|$. One can see that the experimental points in all the three energies for both Totem and Atlas present the shallow zero $|t_{\xi_2}|$ close to 0.05 GeV$^2$ which slowly approaches the origin with increasing energy. However, in the region where the first zero should occur all the data sets present larger errors and/or strong fluctuations but at Totem 13 TeV it shows a trend precisely at $0.0055$ GeV$^2$, where $|t_{\xi_1}|$ is expected to be observed.}
\label{Totem_Atlas_Zeros}
\end{figure}

\begin{figure}
\includegraphics[ scale=0.63]{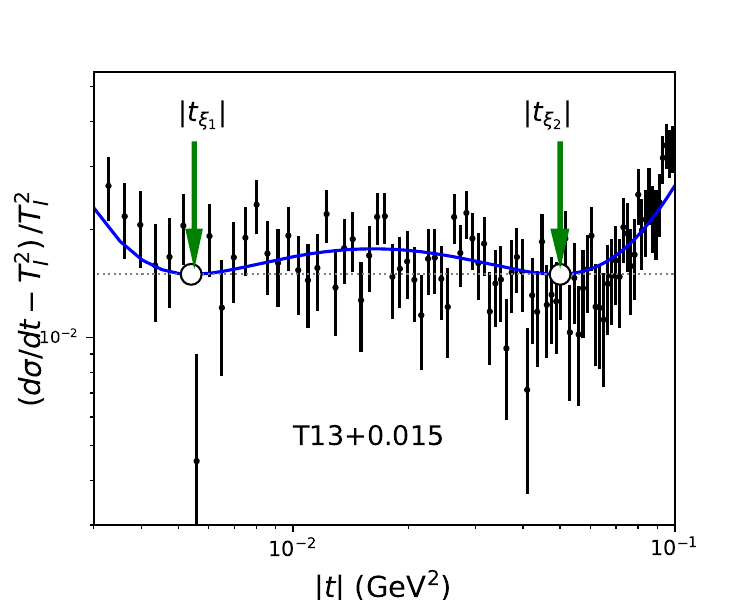}
\caption{Subtracted Totem data at 13 TeV in log-log scale. Since we have negative values when subtract $T_I^N$ from Totem data,  we need to add a factor 0.015 to the subtracted data  in order to make the log-log plot, avoiding negative values. In this representation the trend of the existence of the two zeros is more apparent.}
\label{Totem_close_up}
\end{figure}

In our approach, instead of using a simple exponential reference function, we subtract and then divide the data by the square of the imaginary amplitude, which of course, depends on the chosen model. The interpretation is that the remainder is the square of the sum of the pure real and Coulomb amplitudes divided by the square of the imaginary amplitude
\begin{eqnarray}
\frac{\Big(\frac{d\sigma}{dt}-\pi\,(\hbar c)^2|T_I^N|^2\Big)}{\pi\,(\hbar c)^2|T_I^N|^2}=\frac{|T_R^N+T_C|^2}{|T_I^N|^2}~.
    \label{subtracted_data}
\end{eqnarray}
In Fig. \ref{Totem_Atlas_Zeros} we show the Totem \cite{TOTEM:7TeV,  TOTEM:8TeV,  TOTEM:2017asr}  and Atlas \cite{Atlas:2014,  Atlas:2016, ATLAS:2022}  LHC data subtracted as in Eq.(\ref{subtracted_data}), using for $T_I^N$  the model in Ref.\cite{Kohara:2019qoq} pointing to the possibility of observing the two zeros. The first zero $|t_{\xi_1}|$ is more subtle to observe since it happens in a region where the magnitude of the Coulomb amplitude decreases fast with $|t|$ compared to the nuclear real, which means that this zero should be sharp and  less model dependent, and only more precise experimental data would clearly show this phenomenon. Unfortunately all LHC experiments, except Totem at 13 TeV, have large error bars and/or strong fluctuations in the region where this zero could be observed. In the Totem experimental data at 13 TeV one could see a trend of this dip being formed in Fig. \ref{Totem_close_up} at $|t|\simeq 0.0055$ GeV$^{2}$, which is exactly the position of expected for $|t_{\xi_1}|$. On the other hand the second zero at $|t_{\xi_2}|$ can be clearly seen in all experiments when the non-constant curvature touches zero. However, since in this region the difference in the slopes of the pure real and Coulomb amplitudes is smaller, the behavior of the zero is shallow and its position is more model dependent. We expect that elastic scattering models with explicit real and imaginary amplitudes valid for a broad $t$ range (beyond the diffractive dip) should also present two zeros. 

To summarize, from the experimental point of view, we believe that with large statistics and  better resolution in the very forward region one could observe the first zero $|t_{\xi_1}|$ constraining even more the nuclear real and the Coulomb amplitudes and also the role played by Coulomb phase. Besides, as advocated by Donnachie and Landshoff  \cite{Donnachie_2022}, in the present TOTEM analysis at 13 TeV, the Coulomb phase is forcing the $\rho$ values to be rather small and according to their model the presence of the Coulomb phase will have a negative impact in the description of lower energies.  
In our previous work, we showed that at 13 TeV the  relative phase reduces the value of the parameter $\rho$ and as a consequence the magnitude of the real nuclear part near the origin becomes smaller than the magnitude of the Coulomb part mitigating the existence of first zero \cite{Kohara_2019}. A similar feature was recently noticed by Selyugin \cite{Oleg:2022}, who observed that a smaller value for $\rho$ leads to wrong determination of $t$ dependence of the real amplitude. The existence of $|t_{\xi_1}|$ may be important in the determination of the forward parameter $\rho$ since it is determined in a region where there is a strong interplay between $T_R^N$ and $T_C$. From the point of view of the models, it would be interesting to see how their real part are as compared with the subtracted data. This could be an important test to constrain their real nuclear amplitudes.

\section{Acknowledgments}
  The author warmly thanks Erasmo Ferreira and Takeshi Kodama for encoraging this work, for the insightful comments, suggestions and discussions. The author also thanks Piotr Kotko, Hiren Kakkad and Rafał Staszewski for interesting discussions and inspiring comments.
  The author is supported  by the National Science Centre in Poland, grant no. 2020/37/K/ST2/02665. The research leading to these results has received funding from the Norwegian Financial Mechanism 2014-2021.

\end{document}